# Organization of Historical Oceanic Overturnings on Cross-Sphere Climate Signals

Yingjing Jiang[1], Shaoqing Zhang[1*], Yang Gao[2*], Lixin Wu[1,3*], Lv Lu[1], Zikuan Lin[1], Wenju Cai[1,3], Deliang Chen[4,5], L. Ruby Leung[6], Bin Wang[7], Xueshun Shen[8], Mingkui Li[1], Xiaolin Yu[1], and Xiaopei Lin[1]

[1]Key Laboratory of Physical Oceanography, Ministry of Education/Frontiers Science Center for Deep Ocean Multispheres and Earth System/College of Oceanic and Atmospheric Sciences, Ocean University of China, Qingdao, China

[2]Frontiers Science Center for Deep Ocean Multispheres and Earth System/Key Laboratory of Marine Environmental Science and Ecology, Ministry of Education, Ocean University of China, Qingdao, China

[3]Laoshan Laboratory, Qingdao, China

[4]Department of Earth System Sciences, Tsinghua University, Beijing, China

[5]Department of Earth Sciences, University of Gothenburg, Gothenburg, Sweden

[6]Atmospheric, Climate, & Earth Sciences Division, Pacific Northwest National Laboratory, Richland, WA, USA

[7]Department of Atmospheric Sciences and International Pacific Research Center, University of Hawaii at Manoa, Honolulu, HI, USA

[8]The Center of Earth System Modeling and Prediction, CMA, Beijing, China

**\*Corresponding authors:**

Shaoqing Zhang (szhang@ouc.edu.cn), Yang Gao (yanggao@ouc.edu.cn), Lixin Wu (lxwu@ouc.edu.cn)

**Abstract**

The global ocean meridional overturning circulation (GMOC) is central for ocean transport and climate variations [1–4]. However, a comprehensive picture of its historical mean state and variability remains vague due to limitations in modelling and observing systems [4,5]. Incorporating observations into models offers a viable approach to reconstructing climate history, yet achieving coherent estimates of GMOC has proven challenging due to difficulties in harmonizing ocean stratification [6,7]. Here, we demonstrate that applying multiscale data assimilation scheme that integrates atmospheric and oceanic observations into multiple coupled models in a dynamically consistent way, the global ocean currents and GMOC over the past 80 years are retrieved. While the major historic events are printed in variability of the rebuilt GMOC, the timeseries of multisphere 3-dimensional physical variables representing the realistic historical evolution enable us to advance understanding of mechanisms of climate signal propagation cross spheres and give birth to Artificial Intelligence coupled big models, thus advancing the Earth science.

**Main**

Current understanding of the GMOC that serves as a critical climate component in globally redistributing heat, freshwater, carbon, and nutrients [1–3,8,9] remains restricted because of the lack of its direct observations, necessitating estimation through model-data integration via advanced data assimilation [3,10]. However, given limited deep-ocean observations [6], GMOC estimation is highly challenging since it requires a high-order coherence in ocean vertical structure (i.e. stratification) [7,10] and balance between atmospheric and oceanic variables. Achieving such coherence and balance necessitates both dynamic consistence of multi-spheric as well as multiscale interactions [11,12] as observations are assimilated into models.

Coupled data assimilation (CDA) [13–15] incorporates Earth observational information into coupled model dynamics and physics to rebuild an Earth system close to the real world, aiming to produce dynamically-consistent estimates for three-dimensional (3-D) atmospheric and oceanic motions [16,17], i.e., coupled reanalysis. High-quality coupled reanalysis datasets not only serve as the cornerstone for data-driven artificial intelligence (AI) climate studies (18,19), but also constitute an indispensable foundational resource for giving birth to coupled AI big models (20–22). However, CDA struggles to resolve issues related to insufficient multiscale interactions [23] and the negative effects of systematic model error (i.e. model bias) at deep ocean [24] on assimilation. These challenges lead to divergent estimates of GMOC in existing ocean reanalyses, where mean state and variability often differ substantially [3,25]. The CDA development over decades [13,16,26–30] has advancements in mitigating these issues and CDA is expected to produce useful and reliable estimates of 3-D ocean currents and GMOC [31].

Previously, we have applied a cross-sphere CDA scheme that has multiscale observational constraints [28] and refined deep ocean bias treatments [10] to two leading models used by the Intergovernmental Panel on Climate Change (IPCC): the Community Earth System Model (CESM) [32] and the 2$^{nd}$ generation Coupled Model (CM2) [33], establishing two comprehensive CDA systems [29,30], CESM-CDA and CM2-CDA. Here, we use these two CDA systems to assimilate atmospheric and oceanic observations since 1945, creating two coupled reanalyses spanning over the past 80 years (see **Methods**). Through decomposition and composite analyses of different scale modes on geostrophic and non-geostrophic components, we demonstrate the convergent mean state and variability of the CDA-estimated GMOCs consistent with available "observations." We show that the CDA-estimated atmospheric and oceanic states exhibit coherent configuration (see **Text S1**, **S2** and **Figs. S1–S7**). By analyzing GMOC multiscale components gained in CDA, we also show the physical connections between tropical activities which have very active air-sea interactions where CDA is vital for incorporating data into coupled models, and multidecadal oscillations in high latitudes, which carry the fingerprint of historic events.

**The convergent GMOC mean state in two coupled reanalyses**

We first calculate the mean states of GMOC in two historical model simulations CESM-HIS and CM2-HIS (**Fig.1a&b**), two CDA estimates CESM-CDA and CM2-CDA (**Fig.1c&d**), as well as two ocean reanalysis products SODA3 and ORAS5 (**Fig.1e&f**). In both model-simulated GMOCs, substantial northward transport circulation is observed above 3000 m, underlain by weak southward return circulation. This northward transport circulation features two strong centers: one at 40°–70° S, associated with the Antarctic Circumpolar Current (ACC) system, and the other at 30°–50° N, associated with the North Atlantic Deep Water (NADW) system. Both centers are enclosed within the 10 Sv contour circle. However, substantial differences exist between CESM-HIS and CM2-HIS. For example, the CESM-simulated ACC- and NADW-associated transport centers are stronger than those in CM2; the centers in CM2 are enclosed by a continuous 15 Sv contour, while in CESM, this close 15 Sv contour breaks between 20° and 40° S, resulting in two distinct circulation centers.

In the two CDA estimates, the upper northward transport circulation adjusts to feature three separate centers within the 10 Sv contour circle. The ACC-associated transport becomes weaker, while the NADW-associated transport largely strengthens. Both CDA estimates reveal a new circulation center between 10° and 30° S at a depth of 500–2000 m, while this center appears very weak in SODA3, and absent in ORAS5. The maximum strength of NADW-associated transport in CDA estimates is approximately 20% stronger than in model simulations, extending vertically by 1000 m to reach a depth of 4000 m. Another prominent change in the CDA-estimated GMOCs compared with those of model simulations is intensified deep southward return circulation below 2500 m, exhibiting double centers between 3000 and 4000 m, one at 30° S and the other at the equator. In the reference ocean reanalysis products, the pattern of GMOC mean state resembles a blend of model simulation and CDA estimate: although the upper northward transport circulation exhibits two centers marked by a stronger ACC and weaker NADW within a continuous 10 Sv contour circle, the reanalysis products exhibit a much stronger and wider scope deep southward return circulation than the model simulations and CDA estimates do.

Basin-based decomposition analysis of meridional overturning reveals that the new CDA GMOC center between 10° and 30° S at a depth of 500–2000 m mainly arises from the CDA-enhanced Atlantic subtropical gyre system, resulting in a markedly different Atlantic meridional overturning circulation (AMOC) (**Fig.S6a–g**). In contrast, the substantial southward transport in the CDA estimates largely originates from an enhanced Indo-Pacific deep ocean tropical system, corresponding to a strong Indo-Pacific meridional overturning circulation (IPMOC) (**Fig.S6h–n**). Further detailed analyses of CDA-estimated ocean currents and the GMOC mean state in the **Text S2** support these findings.

**Coherent ocean stratification in balanced coupled estimation**

The convergence of CDA-estimated GMOC mean states can be attributed to the convergent ocean stratification which is consistent with the observation (**Fig.1g–n**). The multiscale CDA constraints greatly reduce the model simulation biases and produce coherent ocean vertical structure, achieving better mean current velocities than the reference ocean reanalyses (**Extend Data Fig.1** and **Fig.S7**). Both CDA estimates align the temperature and salinity structure of the oceans closely with observations, producing ocean stratification that is highly consistent with observational products. Notably, below 2000 m, the reduction of potential density errors of CESM-CDA from CESM-HIS is approximately 60%–90% (**Fig.1g**), and the error reduction of CM2-CDA from CM2-HIS even reaches 95% (**Fig.1h**). Compared with the reference ocean reanalysis products, the ocean stratification of potential densities in CDA estimates demonstrates significant improvements, especially below 2000 m.

This convergence aligns with the observational analysis of 950–1150 m ocean currents based on drifting information from the array for real-time geostrophic oceanography (Argo) floats [34]. Owing to the responses of ocean vertical structures enforced by local data constraints to air-sea interactions (**Figs.S2&S3**), the resultant coherent ocean stratification of CDA adjusts the deep ocean currents and provides a consistent enhancement in the ACC, WBCs, and tropical systems, etc.

**Historical GMOC variability and its physical modes**

The convergent variability of the CDA-estimated GMOCs (see **Methods**) is verifying through volume transports at 47°, 26°, and 16° N in the Atlantic (where long-term observations exist), as well as the geostrophic components against the observational dataset EN4 [35] (**Extend Data Fig.2** and **Text S2g**). To understand the mechanism of CDA in reproducing GMOC variability, we decompose the multiscale components of model-simulated and CDA-estimated GMOCs. The GMOC theory [36] suggests the existence of multiscale processes that balance atmospheric forcing and oceanic response to sustain the GMOC. In CDA, atmospheric data constraints refine the wind system and correct the atmospheric mean meridional circulation (AMMC), which drives ocean circulation, while data-constrained ocean currents improve atmospheric feedback through refined heat fluxes (**Fig.S2&S3**). The balance between the residual of Southern Ocean upwelling minus Northern Hemisphere sinking and the low-latitude diffusive mixing maintains the AMMC-GMOC structure [2,37] (illustrated in **Extend Data Fig.3**). The NADW system represents mostly the Northern Hemisphere sinking [38]. The Southern Ocean upwelling is represented by the residual circulation (RC) from the wind-driven Ekman effect and eddy-induced transport on the flanks of submarine ridges along the ACC [39], i.e., the ACC-RC system. And we calculate the kinetic energy dissipation rate of 30° S–30° N, $\varepsilon_m$, to examine tropical diffusive mixing effects [40] (see **Methods**). For the mean state, NADW estimated by both CDAs produces a nearly identical isopycnal structure with the observation while ORAS5 only displays somewhat improvement from free model simulations (**Fig.1g–l**). The CDA-estimated ACC-RCs also appear closer to the "observation" (**Fig.S11**), while the $\varepsilon_m$ estimates are close to the level of high-resolution (25 km) ORAS5 ocean reanalysis (**Fig.S12**).

We perform power spectrum analysis and examine band-pass filtered timeseries of GMOC, NADW, ACC-RC, and $\varepsilon_m$ indices on decadal (5–20-yr) (**Fig.2**), interannual (2–5-yr) (**Extend Data Fig.4**) and seasonal-to-interannual (6-month–2-yr) (**Extend Data Fig.5**) scales. We find that the anomaly correlation between two model-simulated GMOCs is very small in these three scale bands (0.18, 0.22, and 0.45 respectively), whereas the two CDA-estimated GMOCs are significantly correlated in all scale bands (0.62, 0.73, and 0.89 respectively). This means that the CDA model-data integration does capture multiscale information of the Earth system to some degree. In general, the NADW and ACC-RC decadal modes in all data-constrained products highly correlate with the "observation" in which the correlations of NADW and ACC-RC in ORAS5, CESM-CDA, and CM2-CDA by order are 0.60, 0.93, 0.86 and 0.70, 0.68, 0.58, respectively (**Fig.2b&c**). For the seasonal to interannual scales, while the ACC-RC highly correlates with the "observation" (the correlations in ORAS5, CESM-CDA, and CM2-CDA by order are 0.65, 0.73, 0.79 in the interannual band and 0.82, 0.86, 0.73 in the seasonal-to-interannual band), the NADW in all data-constrained products has however no significant correlation with the "observation" (**Extend Data Fig.4b&5b**). This suggests that while the CDA is able to coherently integrate data into coupled models and rebuild the 3-D geo-fluids, it is also constrained by the model's deficiency. Because of absent mesoscales in the 100 km coarse resolution assimilation models used in this study, it's very difficult for the CDA to retrieve high-frequency signals of NADW, for which mesoscale eddies play an important role [41].

We also find that while the GMOC's decadal oscillations starting from the early 1980s has a significant fingerprint on the decadal modes of NADW, ACC-RC, and $\varepsilon_m$ in both CDA estimates (**Fig.2b–d**). In the seasonal to interannual scales, the GMOC significantly correlates with the corresponding ACC-RC mode (**Extend Data Fig.4c&5c**). At the interannual band, the correlations of GMOC with ACC-RC in ORAS5, CESM-CDA, and CM2-CDA are 0.61, 0.74, and 0.69, respectively, and at the seasonal-to-interannual band, the correlations are 0.33, 0.65, 0.37 respectively, whereas NADW and $\varepsilon_m$ have no robust correlation with GMOC in these bands. This phenomenon could be attributed to the multiscale characteristics of ACC-RC as a strongly-coupled atmosphere-ocean component of the Earth system, whereas the seasonal to interannual scale NADW and $\varepsilon_m$ modes strongly rely on oceanic mesoscale activities and associated air-sea interactions [42], which are unresolved in the models used in this study.

To better understand the key processes that maintain GMOC, we perform a composite analysis of GMOC variability in terms of the NADW, ACC-RC, and $\varepsilon_m$ (**Fig.3** and **Extend Data Fig.6**). We first calculate the first principal component of GMOC stream functions over regions in the north of 30° N, between 30° S and 30° N, as well as in the south of 30° S (denoted as $MOC^n$, $MOC^t$ and $MOC^s$). We find that the $MOC^n$, $MOC^t$, and $MOC^s$ generally correlate with the NADW, $\varepsilon_m$ and ACC-RC indices well (detailed analyses are provided in **Text S3c**). Variability of the GMOC is largely accounted for by the linear combination of ACC-RC, NADW, and $\varepsilon_m$ in both free model simulations and data-constrained products, and such fit is better in free model simulations. The mean correlation between the original and synthesized GMOC indices is 0.69 for the two free model cases and 0.45 for the three data-constrained cases. This reflects how well models tend to express idealized relationships between variables, while data constraints inevitably introduce some imbalances as model misfits are corrected by data. However, the correlation in the two CDA estimates is much higher (0.55) than in the ORAS5 ocean reanalysis (0.31), suggesting that the approach of CDA data constraint may help reduce imbalances introduced during data-model integration.

**GMOC's fingerprints on historic events**

Previous model studies have shown that if superimposed on a cold phase of surface ocean internal variability, the cooling effect of natural aerosols caused by a volcanic eruption can greatly enhance and extend the cold abnormality of sea surface temperature (SST) in a global scope [43]. Two CDA estimates produce nearly identical global mean SST residual (GMSSTr) (a quadratic fit of GMSST is removed [43]) variability on Mt. Agung (1963), Mt. Chichon (1982) and Mt. Pinatubo (1991) events (**Fig.S15a**), which shows the same features of observed volcanic eruptions as been detected [43]. Here, we detect the consequence of such cold SST anomalies caused by these historic volcanic eruptions impacting NADW, so as on GMOC, ACC-RC, and $\varepsilon_m$. We find that the SST anomaly (SSTA) distributions produced by the Chichon and Pinatubo volcanic eruptions are very similar, whereas the distribution produced by the Agung is very different (**Fig.4a–c**). While the Agung volcanic eruption produced cold SSTAs in a large area of the Southern Ocean and warm SSTAs in the North Atlantic, the Chichon and Pinatubo volcanic eruptions mainly produced cold SSTAs in the North Atlantic. To detect possible linkage between the volcanic eruption-caused SSTA and the NADW signal, we examine the timeseries of North Atlantic SST residual (NASSTr) and NADW indices (**Fig.4d**). We find that the NADW's variations have generally an opposite phase with the NASSTr by a roughly 2–3-yr lag (**Fig.4e&f**), reflecting a speed of a couple of meters per day for local water vertical transport in the North Atlantic [44]. Due to enhancement by the cooling effects of the Chichon and Pinatubo events, the NASSTr's variations appear as a very strong intra-decadal cold phase from the early 1980s to the end of 1990s, with its bottom in 1993–1994. The surface cold water corresponds to the strong NADW phase over 1990–2002, with its peak in 1996–1997.

The AMMC anomalies (departure from time mean) caused by the Agung and Chichon/Pinatubo events are quite different (**Fig.4g&h**), with the Agung-caused anomalies being large in the Southern Hemisphere while the Chichon/Pinatubo-caused anomalies reside mainly in tropics and high-latitudes of the Northern Hemisphere. This suggests that the atmospheric background wind systems during the Agung and Chichon/Pinatubo are quite different, resulting in different aerosol distributions driving different anomalous atmospheric circulations. Indeed, a strong positive phase of the North Atlantic Oscillation dominates the anomalies of atmospheric circulations during the Chichon/Pinatubo events, while during the Agung event, the Southern Annular Mode dominates the atmospheric circulation anomalies (**Fig.S18**).

**Discussions**

The multiscale characteristics of GMOC and its components such as NADW, ACC-RC, and decadal tropical kinetic energy dissipation rates are analyzed in detail using scale decomposition and composite analyses. While the GMOC's fingerprints on historic events have been identified, accurately reconstructed 3-D ocean currents and GMOC are expected to further provide key insights into the sources of predictability for extreme climate phenomena associated with internal modes of the coupled system. With the multiscale representation provided by coupled reanalysis data, advanced artificial intelligence and machine-learning forecasting methods [45,46] can be further developed to enhance seamless prediction capabilities. First, coupled reanalyses with models that can resolve tropical cyclones, ocean mesoscale eddies, clouds, and sub-mesoscales [47] are essential to boost the multiscale representation of the Earth system. Second, sea-ice genesis and melting are inherently multiscale phenomena, encompassing scales ranging from millimeters to thousands of kilometers, each playing a uniquely role [48]. This multiscale complexity poses specific challenges to high-resolution sea-ice modeling in both thermodynamic processes and rheological dynamics [48,49]. Given the booming availability of multisource multisphere data, refined coupled reanalysis that incorporates such data into high-resolution models with deep machine-learning schemes shall provide a plausible way to break such bottlenecks.

**Figs. 1–4:**

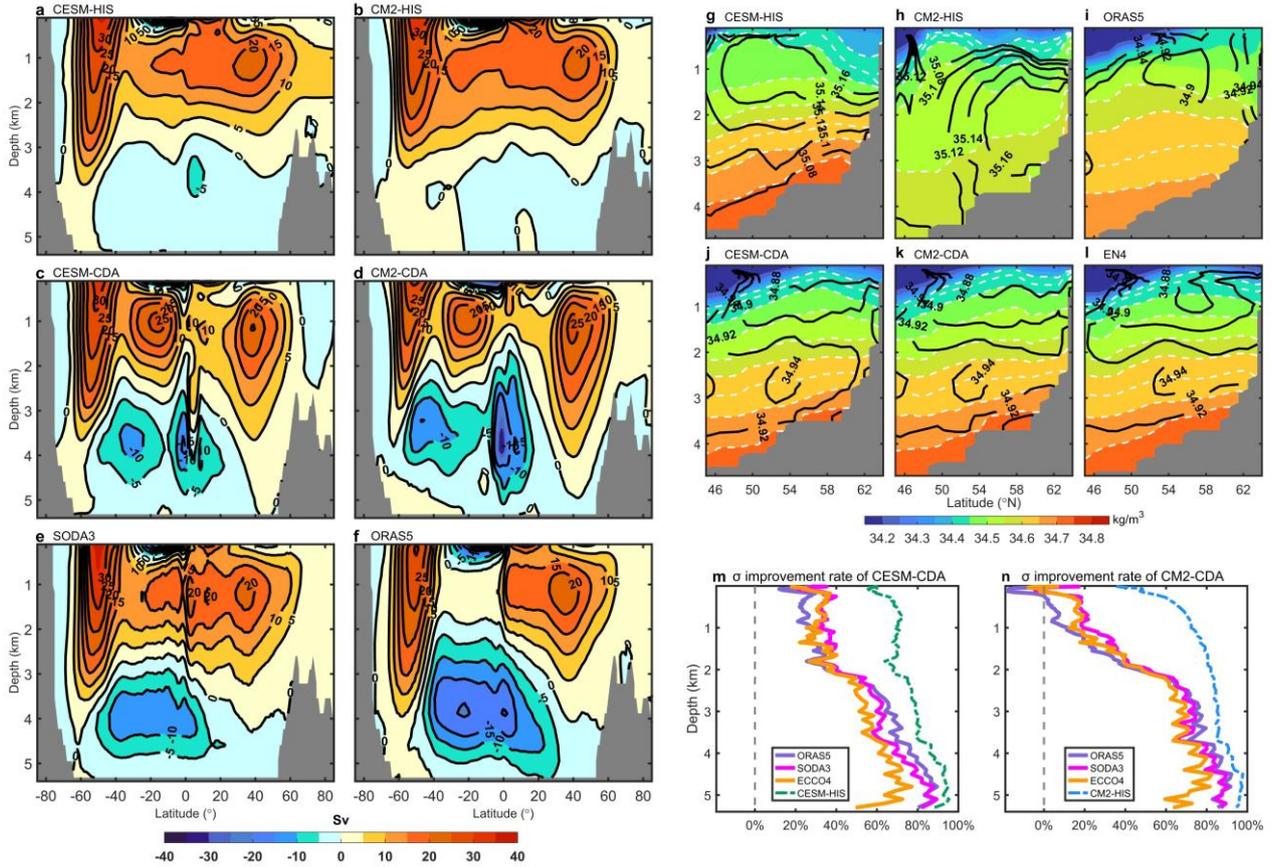

**Fig.1 | The convergent mean state of global meridional overturning circulations (GMOCs) estimated in two coupled reanalyses. a–f**, Stream functions (unit: Sv = $10^6$ m$^3$/s) of GMOCs for historical simulations (1950–2022) of CESM and CM2 (denoted as CESM-HIS and CM2-HIS), coupled data assimilation (CDA) results (1950–2022) with CESM and CM2 models (denoted as CESM-CDA and CM2-CDA), as well as the ocean reanalysis products SODA3 (1980–2019) and ORAS5 (1958–2022). **g–l**, Mean states of potential density distributions in the North Atlantic over the domain of 55°–35° W for CESM-HIS (**g**), CM2-HIS (**h**), ORAS5 (**i**), CESM-CDA (**j**), CM2-CDA (**k**) and observational dataset EN4 (**l**). The shadings with white contours represent potential density referenced to a depth of 1500 m (unit: kg/m$^3$). The black contours represent salinity (unit: PSU). **m–n**, Potential density (σ) root mean square error (RMSE) reduction rate against the observation product WOA18 for CESM-CDA and CM2-CDA compared with their respective model simulation as well as reanalysis products.

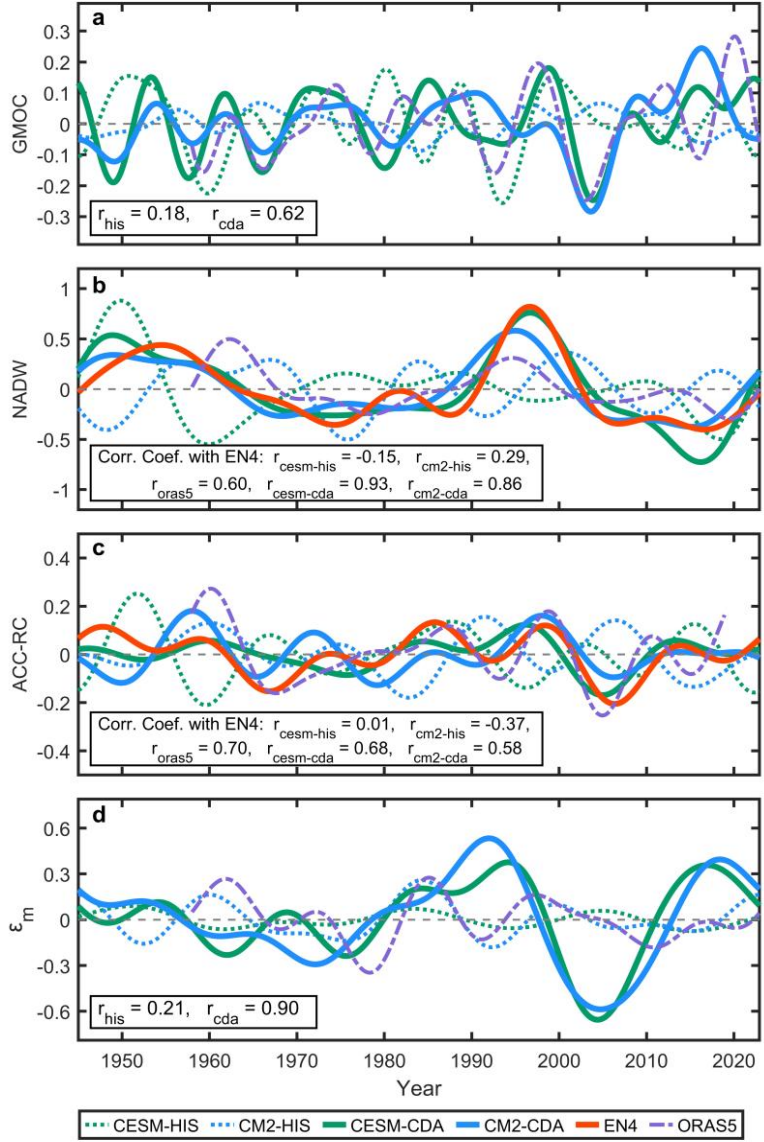

**Fig.2 | Decadal variability of the GMOC and associated physical modes in coupled reanalyses. a–d**, 5–20 years filtering of time series for GMOC (**a**), North Atlantic Deep Water (NADW) (**b**), Antarctic Circumpolar Current residual circulation (ACC-RC) (**c**) and tropical diffusive mixing ($\varepsilon_m$) indices (**d**). The correlation coefficients between various curves and observational products are listed within a box in **b** and **c**. The correlation coefficients between the two model simulations and those between the two CDA estimates are listed within a box in **a** and **d**. The calculation periods of all correlation coefficients are consistent with those of ORAS5.

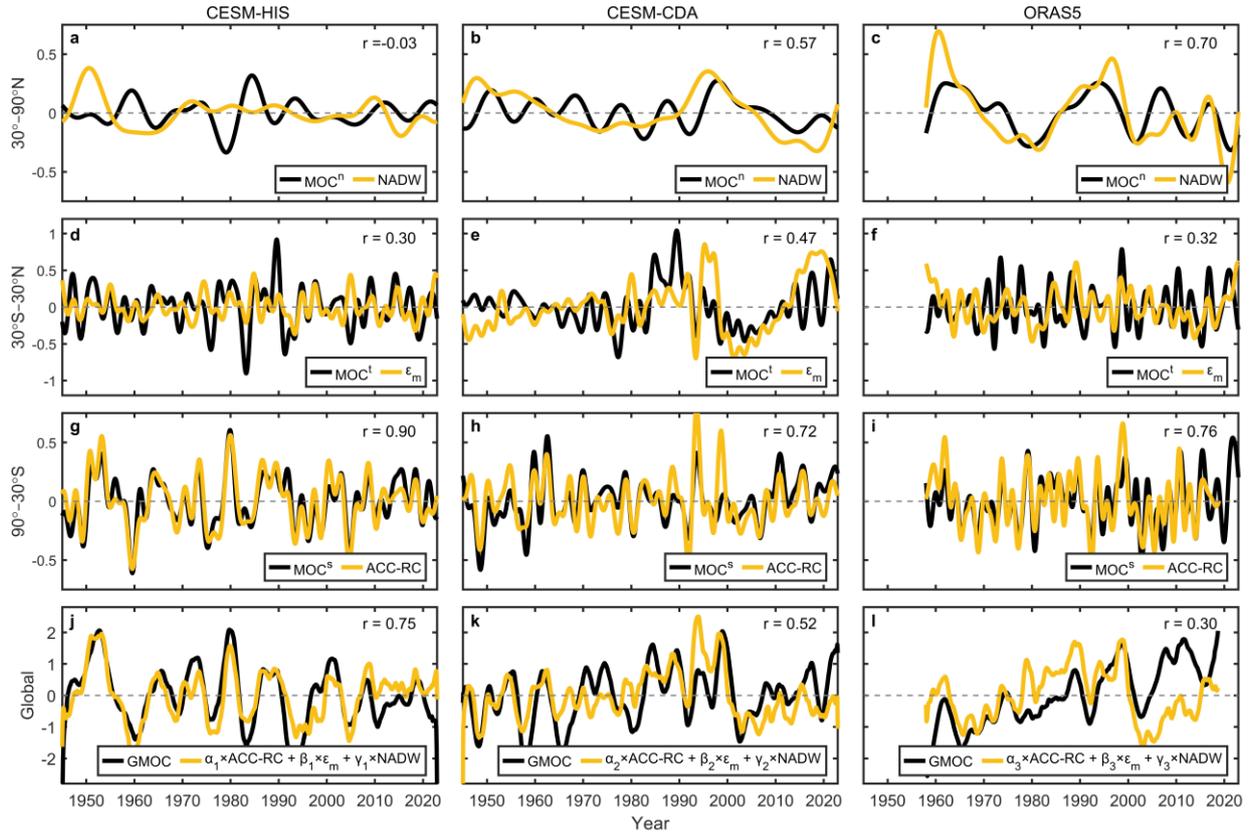

**Fig.3 | Composited GMOC with its physical modes in CESM model simulation and coupled reanalysis. a–c**, The first principal component (PC1) derived from empirical orthogonal function (EOF) decomposition of GMOC over the area north of 30° N (denoted as MOC$^n$) and 5–30-yr filtered time series of normalized North Atlantic Deep Water (NADW) for CESM-HIS (**a**), CESM-CDA (**b**), and ORAS5 (**c**). **d–f**, Same as **a–c**, but for PC1 of GMOC over the area of 30° S–30° N (denoted as MOC$^t$) and 6-month–20-yr filtered kinetic energy dissipation rate ($\varepsilon_m$). **g–i**, Same as **a–c**, but for PC1 of GMOC over the area south of 65° S (denoted as MOC$^s$) and the 1–20-yr filtered Antarctic Circumpolar Current residual circulation (ACC-RC). **j–l**, Time series of the 24-month running mean GMOC index and linear combination of ACC-RC, $\varepsilon_m$, and NADW with coefficients α, β, and γ, respectively. The parameters α, β, and γ are the linear regression coefficients between the GMOC indices and ACC-RC, $\varepsilon_m$, NADW indices, respectively, as [$\alpha_1$, $\beta_1$, $\gamma_1$] = [0.64, −0.05, 0.24], [$\alpha_2$, $\beta_2$, $\gamma_2$] = [0.46, 0.12, 0.07], and [$\alpha_3$, $\beta_3$, $\gamma_3$] = [0.31, −0.01, 0.08].

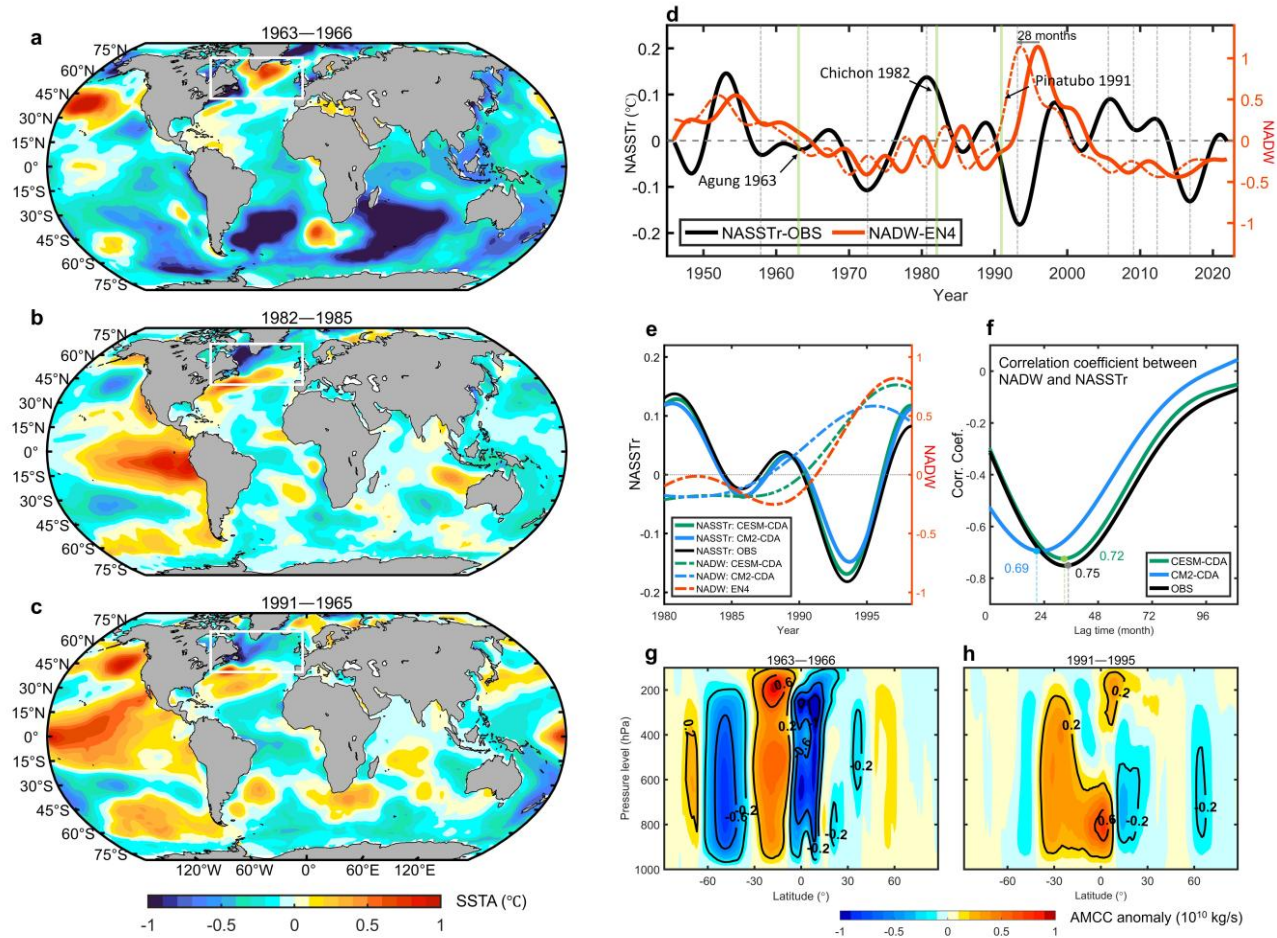

**Fig.4 | The response of NADW on major volcanic eruptions in the 2nd half of 20th Century. a–c**, Distribution of sea surface temperature anomaly (SSTA) during the periods of 1963–1965 (**a**), 1982–1985 (**b**) and 1991–1995 (**c**) for observation (OBS). These three global SST cooling periods correspond respectively to the three major volcanic eruptions of the second half of the 20th Century, namely Mt. Agung (1963), Mt. El Chichon (1982), and Mt. Pinatubo (1991) [43]. **d**, Time series of 5–20-yr band-pass filtering NASSTr (see **Text S3d** for the definition) for OBS (black-solid, °C) and 2–20-yr band-pass filtering normalized North Atlantic Deep Water (NADW) for EN4 (red-solid). The times of the three volcanic eruptions events are marked by light-green straight lines. The red-dashed curve is obtained by shifting the red-solid curve backward by 28 months and each dashed-grey line marks an opposite phase between the lagged-NADW and NASSTr. **e**, 5–20-yr band-pass filtering NASSTr and NADW during 1980–2000. **f**, Lag-correlation relationship between NADW and NASSTr in **e**. **g–h**, Atmospheric mean meridional circulation (AMMC) anomaly distributions during 1963–1965 (**g**) and 1911–1995 (**h**).

**Methods**

**Coupled data assimilation systems**

**Two coupled models.** The CESM [32] and CM2 [33] employed the historical simulation period from 1945 to 2006 and the Representative Concentration Pathway 4.5 (RCP45) scenario period from 2007 to the present. For CESM, the atmospheric component (Community Atmosphere Model version 5, CAM5) employed a spectral-element dynamic core ("ne30g16" configuration) with approximately 1° × 1° horizontal resolution near the equator, and 26 vertical levels; the oceanic component (Parallel Ocean Program version 2, POP2) is configured with a horizontal resolution of approximately 1° × 1° and 60 vertical levels, with 15 levels of 10-m thickness in the upper 150 m. For CM2, the atmospheric component (Atmosphere Model version 2.1, AM2.1) employed a finite-volume dynamic core with a horizontal resolution of 2° latitude × 2.5° longitude and 24 vertical levels; the oceanic component (Modular Ocean Model version 4, MOM4) is configured with 50 vertical levels (22 of which are 10-m thick in the upper 220 m) and a 1° × 1° horizontal B-grid resolution, telescoping to 1/3° meridional spacing near the equator.

**Multiscale data assimilation algorithm.** A multi-timescale, high-efficiency approximate Ensemble Kalman Filter [28], combined with a deep-ocean bias relaxation scheme [10], is implemented into both models to establish two CDA systems. This CDA algorithm uses the single-model state time series to construct stationary, low-frequency, and high-frequency filters, markedly reducing the consumption of computational resources [29,30]. The algorithm consists of the following two steps:

$$\Delta y^{o,(\tau)} = \frac{\frac{1}{(\sigma^{(\tau)})^2} y_f^p + \frac{1}{(\sigma^{(o)})^2} y^o}{\frac{1}{(\sigma^{(\tau)})^2} + \frac{1}{(\sigma^{(o)})^2}} - y_f^p, \quad \tau = 1, \quad \Gamma \ (\Gamma = 3 \ in \ this \ study), \quad (1)$$

$$\Delta x^u = \sum_{\tau=1}^{\Gamma} \alpha^{(\tau)} \frac{cov(\Delta x^{(\tau)}, \Delta y^{(\tau)})}{(\sigma^{(\tau)})^2} \cdot \Delta y^{o,(\tau)}, \quad \tau = 1, \quad \Gamma \ (\Gamma = 3 \ in \ this \ study). \tag{2}$$

Here, $\Delta y^{o,(\tau)}$ represents the observational increment, $y_f^p$ is the "prior" model-estimated value at the observation location, $\sigma^{(\tau)}$ is the standard deviation of the multi-timescale (three timescales: stationary, low-frequency, and high-frequency in this case) ensemble, $\sigma^{(o)}$ is the standard deviation of the observation, $y^o$ is the observation, $\Delta x^u$ is the observation increment on the model grid after projection, $cov(\Delta x^{(\tau)}, \Delta y^{(\tau)})$ is the covariance relationship between the model value at the model grid and the estimated model value at the observation location of each timescale filtering scheme, and $\alpha^{(\tau)}$ represents the adjustable weight coefficients in the combination of filters.

**Deep-ocean bias treatment.** Given the scarcity of deep-ocean observations, an ocean model bias relaxing scheme [10] is used during CDA to control deep-ocean model bias and achieve coherent ocean stratification [3,25]. In this scheme, climatological temperature and salinity data are restored into the model space with depth-dependent restoration strength. Specifically, the upper ocean is primarily constrained by observations, while the deep ocean is relaxed towards the climatology of the World Ocean Atlas (WOA) [50,51]. On the basis of real-observation assimilation experiments with CM2 [29] and CESM [30], the restoration timescale is linearly related to depth, ranging from infinite to 360 days as depth increases from the surface to 1000 m, then decreasing to nearly 180 days at 1500 m. At greater depths, the restoration strength increases further, with a timescale of 30 days at the ocean bottom.

**Coupled data assimilation scheme.** Ocean observations—including sea surface temperatures (SSTs) from the Hadley Centre Sea Ice and Sea Surface Temperature Dataset (HadISST) and Optimum Interpolation SST (OISST), as well as in-situ ocean temperature and salinity profiles since 1945—are assimilated into the CDA systems. To simplify the CDA framework for climate reanalysis, only surface pressure ($P_s$) data from the fifth-generation ECMWF reanalysis (ERA5) atmospheric reanalysis products [52] are assimilated as the atmospheric "observations" (see **Text S1c**). The $P_s$ data effectively represent all other atmospheric information by capturing whole-column atmospheric mass variation [52]. The assimilation of $P_s$ increments uses an inverted projection method of the vertical integral, effectively extracting observational information into the models [53,54]. The CESM-CDA and CM2-CDA systems are initialized from their coupled states on January 1, 1945, within their respective historical simulations. These systems completed their coupled reanalysis to the present, using historical radiative forcings, and followed the RCP45 scenario setting after 2007 (see **Table S1**).

**Observation-based data and reanalysis products**

To assess GMOC variability, we utilize Atlantic meridional overturning circulation (AMOC) volume transport observations from the RAPID-MOCHA array (Feb. 2004–Dec. 2021) at 26.5° N in the Atlantic [55], along with volume transport data from the Meridional Overturning Variability Experiment (MOVE) mooring observations (Jan. 2000–Jan. 2018) at 16° N in the western Atlantic and the North Atlantic Changes (NOAC) array (Jan. 1993–Dec. 2018) at 47° N [56]. Mid-depth (950–1150 m) ocean current velocities estimated by CDA are evaluated using Argo float displacement data [34] from the Argo New Displacements Rannou and Ollitrault dataset [57]. Climatological data for ocean temperature and salinity from the WOA, as well as monthly EN4.2.2 temperature and salinity datasets (referred to as EN4) [35], are used for further evaluation.

Ocean reanalysis products used for comparison included the Simple Ocean Data Assimilation version 3.4.2 (denoted as SODA3) for the period 1980–2019 [58], Estimating the Circulation and Climate of the Ocean version 4 release 3 (denoted as ECCO4) for the period 1992–2015 [59] from the National Aeronautics and Space Administration, and Ocean Reanalysis System 5 (ORAS5) for the period 1958–2022 [60] from the ECWMF. Additionally, ERA5 atmosphere reanalysis data from 1945 onwards are incorporated. Detailed information on these reanalysis datasets is provided in **Table S2**.

**Definition of GMOC and associated indices**

**Global meridional overturning circulation.** The stream function of GMOC is defined as zonally integrated volume transport, measured in Sv (1 Sv = $10^6$ m$^3$/s), from the depth level z to the ocean surface $\eta$ at latitude y [10,25,30]. It can be expressed as:

$$\Psi(y,z) = \int_z^\eta \int_{x_W}^{x_E} v(x,y,z) dx dz, \quad (3)$$

where, $v$ represents the meridional velocity, and $x_W$ and $x_E$ denote the western and eastern boundaries of the global ocean, respectively.

**GMOC index.** Given that the spatial patterns of the principal component (PC) obtained through empirical orthogonal function (EOF) analysis are not consistent across different models, we employ a novel method to define the GMOC index. The GMOC index is defined as the linear regression coefficient of the instantaneous time slice (monthly in this case) GMOC state over the temporal mean distribution of the CDA mean state. The formula is expressed as:

$$G_{moc} = \frac{cov(G_t, G_m)}{\sigma_{G_t} * \sigma_{G_m}}, \quad (4)$$

where, $cov(G_t, G_m)$ represents the covariance between $G_t$ and $G_m$. Specifically, it is calculated as $cov(G_t, G_m) = \frac{\frac{1}{J_0}\sum_j \frac{1}{K_0}\sum_k G_t(j,k)*G_m(j,k)*\sigma_{G_t}}{\sigma_{G_m}}$ (where j and k denote the indices for the y–z plane, and the sums are taken over the appropriate ranges of j and k). Here, $G_m$ is the mean state of CDA GMOC, $G_t$ is the instantaneous time slice of GMOC, $J_0$ and $K_0$ are numbers of points on the y–z plane, and $\sigma_{G_t}$ and $\sigma_{G_m}$ are the standard deviations of $G_t$ and $G_m$, respectively.

**Geostrophic component of GMOC.** We calculate the geostrophic GMOC using observed and CDA-derived ocean temperature and salinity data, and subsequently verified its geostrophic component. The following formula for the geostrophic meridional velocity ($v_g$) is derived based on the geostrophic balance and thermal wind relationship, which are fundamental in oceanographic studies [61]:

$$\boldsymbol{v_g}(k) = \boldsymbol{v_g}(k+1) - \frac{g}{f}\frac{[z(k+1) - z(k)]}{\sigma(k)}\frac{\partial \sigma(k)}{\partial x}, \qquad (5)$$

where, $g$ is the gravitational acceleration, $f$ is the Coriolis parameter, $k$ is the *k-th* vertical grid with $k$ increasing downwards, $z$ is the grid depth, and $\sigma(k)$ is the potential density referenced to $z(k)$. The geostrophic velocity is reconstructed with the level of no motion at the ocean bottom. Referring to previous research, a small spatially uniform velocity *O (0.1 mm/s)* is removed afterwards to ensure zero meridional transport [62].

**North Atlantic Deep Water.** The NADW index is defined as the average thickness between two isopycnals of $\sigma_{1.5}$ over the domain of 55°–35° W and 45°–65° N [63,64]. Here, $\sigma_{1.5}$ represents the potential density referenced to a depth of 1500 m. In this study, we chose $\sigma_{1.5}$ = (34.56, 34.68) as the two isopycnals for CESM-HIS, CESM-CDA, CM2-CDA, ORAS5, and EN4, and $\sigma_{1.5}$ = (34.52, 34.56) for CM2-HIS. The differences in the isopycnals chosen for different simulations are attributed to the distinct characteristics of deep mode water formation and evolution in each model.

**Residual circulation of the Antarctic Circumpolar Current.** The ACC-RC denotes the net transport of Ekman effects, while being compensated by eddy-induced transport [38]. It reflects the overall balance between the direct impact of Ekman effects and the counteracting influence of eddy-induced transport. We calculate the ACC-RC from the following formula [65]:

$$\Psi^* = -\frac{\tau}{f} + K_e S, \qquad (6)$$

where, $\tau$ represent wind stress, $f$ is the Coriolis parameter, $K_e$ is eddy diffusivity coefficient, which is set to *1000 $m^2/s$* in this study [66], and $S$ is the slope of isopycnal. Similar to the GMOC index, we define the ACC-RC index as the linear regression coefficient of the instantaneous time-slice (monthly in this particular case) distribution of the ACC-RC system in relation to its temporal mean distribution.

**Tropical diffusive mixing.** We evaluate the kinetic energy dissipation rate for the analysis of the tropical diffusive mixing effects, which are crucial for maintaining global circulation. Because the turbulence is not resolved in our experiments, only the diagnosed kinetic energy dissipation rate of turbulence is computed. The following equation represents the diapycnal advection-diffusion balance, which is fundamental in our analysis:

$$\frac{D\rho}{Dt} = \frac{\partial(\Gamma \varepsilon N^{-2} \frac{\partial \rho}{\partial z})}{\partial z}, \quad (7)$$

where, $N^{-2} = -\frac{g}{\rho_0}\frac{\partial \rho}{\partial z}$, $\rho$ represents the potential density, $\rho_0$ is the reference density, and $\varepsilon$ is the diagnosed kinetic energy dissipation rate. The buoyancy flux is related to the turbulence $\varepsilon$ through the constant efficiency factor $\Gamma$ [67]. The expression for $\varepsilon$ on each layer, which is derived from relevant physical principles, is as follows:

$$\varepsilon = \int_z^0 -\frac{g}{\Gamma \rho_0}\frac{d\rho}{dt}dz + C(x,y). \quad (8)$$

In this study, we set $\Gamma=0.2$ and $\rho_0=1026$ $kg/m^3$. The constant $C$ is ascertained by ensuring that the minimum value of the integration equals 0. The vertical mean of $\varepsilon$ is defined as the mean energy dissipation rate $\varepsilon_m$ in 30° S–30° N. The tropical energy dissipation index $\varepsilon_m(t)$ is defined as the linear regression coefficient of the monthly time slice of $\varepsilon_m$ with respect to its temporal mean distribution.

**Correlation statistics significance**

Correlation significance is assessed via t-test, with critical values ($r_\alpha$) calculated as: $r_\alpha = \frac{t}{\sqrt{t^2+df}}$. For a 960-month sample ($\alpha=0.001$), coefficients $r>0.105$ indicate 99.9% significance, while $r>0.089$ denotes 99% significance. For ORAS5 comparisons (780 monthly data points, 1958–2022), the critical $r_\alpha$ is 0.117.

# Data availability

Data related to the manuscript can be downloaded from the following: ERA5, https://cds.climate.copernicus.eu/datasets; HadISST, https://www.metoffice.gov.uk/hadobs/hadisst/; OISST, https://www.ncdc.noaa.gov/oisst/data-access; Argo, ftp://ftp.ifremer.fr/ifremer/argo/geo/; WOA, http://www.ncei.noaa.gov/; EN 4, https://www.metoffice.gov.uk/hadobs/en4/; SODA3, https://www2.atmos.umd.edu/; ECCO4, https://ecco-group.org/; ORAS5, https://cds.climate.copernicus.eu/; RAPID, https://rapid.ac.uk/; MOVE, https://usclivar.org/amoc/amoc-time-series; NOAC, https://doi.org/10.1594/PANGAEA.959558. Experimental data are available via Zenodo at https://doi.org/10.5281/zenodo.15070241 or from the corresponding author on request.

# Code availability

The code for climate model data processing and analyses is available via Zenodo at https://doi.org/10.5281/zenodo.15070241.

**Acknowledgements**

We thank all researchers who reported and kindly provided us with precious data on to support this study.

**Funding**

This work was supported by the Science and Technology Innovation Project of Laoshan Laboratory (LSKJ202300400, LSKJ202300401–03), the Science and Technology Innovation Project of Laoshan Laboratory (LSKJ202202200, LSKJ202202201–04), and the National Natural Science Foundation of China (42361164616).

**Author contributions**

YJ and SZ designed and led the research and drafted the manuscript. YJ processed the data and drew the figures under the supervision of SZ and LW. YJ, SZ and YG analyzed the data and co-wrote the paper. All other coauthors joined the research and discussions, making comments to the manuscript.

**Competing interests**

The authors declare no competing interests.

**Extend Data Figs.1–6:**

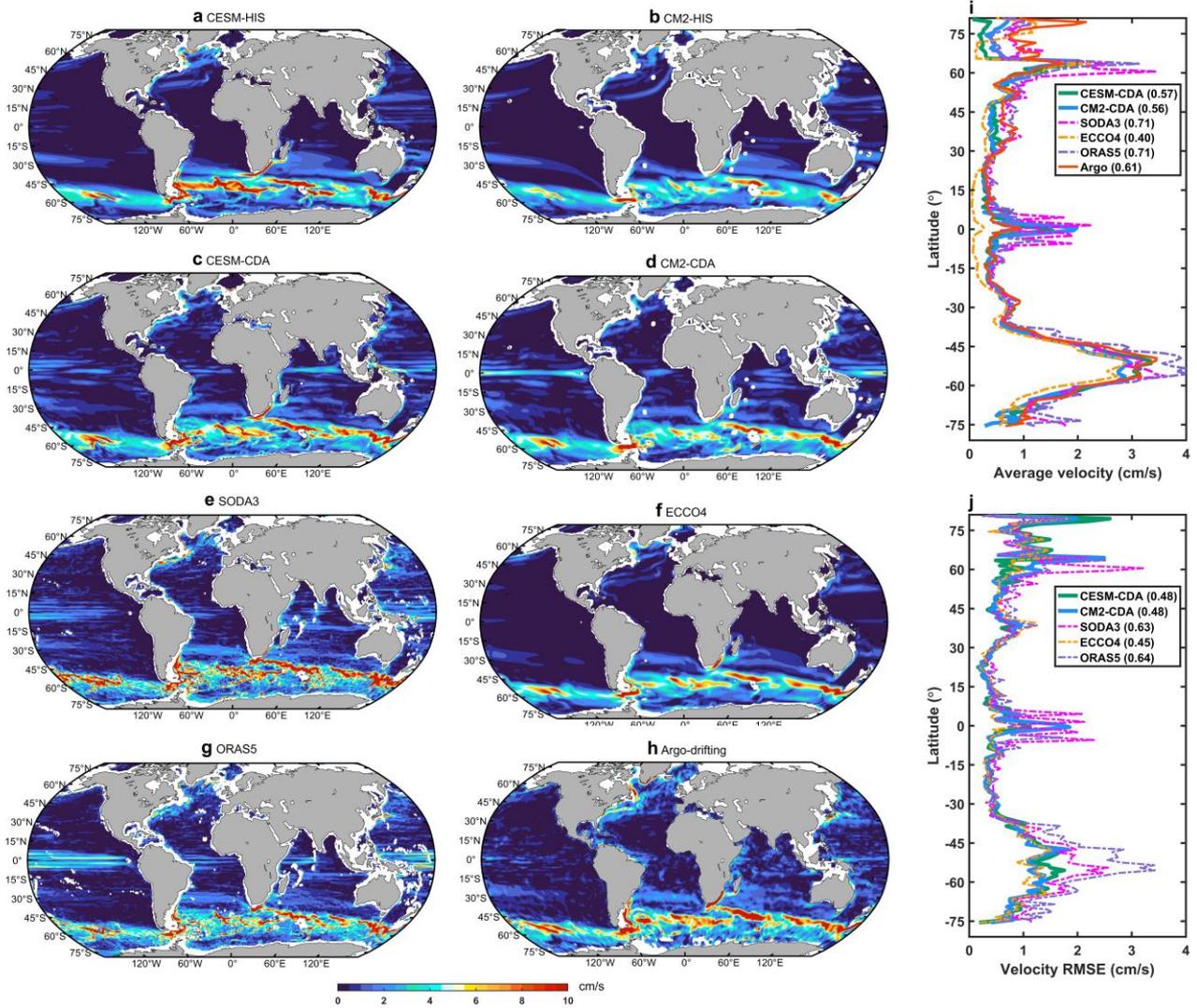

**Extend Data Fig.1 | Verification of ocean currents in coupled reanalyses. a–b**, Argo period (2000–2023) temporal mean distributions of 950–1150 m ocean current velocities (unit: cm/s) in CESM-HIS and CM2-HIS, respectively. **c–g**, Same as **a–b** but for CESM-CDA, CM2-CDA, SODA3, ECCO4 and ORAS5, respectively. **h**, Argo-drifting estimated ocean current velocities (unit: cm/s) for 950–1150 m ocean depth [34]. **i**, The distribution of average ocean current velocities with latitude at 950–1150 m. **j**, Same as panel **i**, but for root mean square errors (RMSEs).

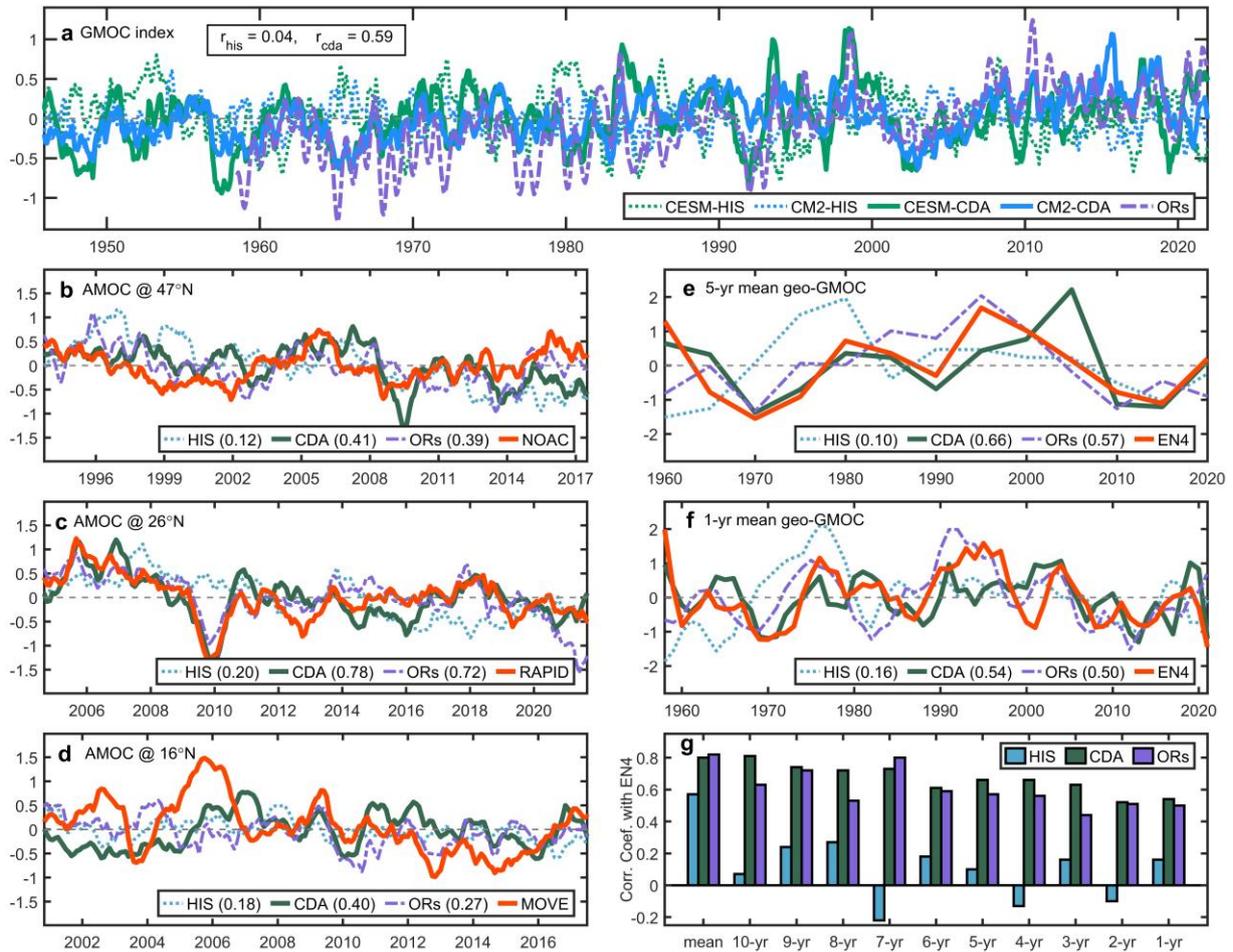

**Extend Data Fig.2 | Observation-convergent variability of estimated GMOCs in coupled reanalyses. a**, Time series of 12-month running mean GMOC indices in two model historical simulations (CESM-HIS and CM2-HIS) and CDA estimates (CESM-CDA and CM2-CDA), as well as the mean of three ocean reanalysis products (denoted as ORs). **b–d**, Time series of meridional mass transport anomalies at 47°, 26°, and 16° N sections in the Atlantic, respectively, for the mean of CESM-HIS and CM2-HIS (denoted as HIS), mean of CESM-CDA and CM2-CDA (denoted as CDA), ORs, as well as the observations of NOAC (**b**), RAPID (**c**), and MOVE (**d**) arrays. **e–f**, Time series of geostrophic GMOC indices with 5-year mean data (**e**), and yearly mean data (**f**), in HIS, CDA, ORs and the ocean objective analysis product EN4. **g**, Variation of anomaly correlation coefficients between the HIS, CDA, and ORs geostrophic GMOCs and the EN4 geostrophic GMOC as the timescale changes from low-frequency to high-frequency. The mean state correlation coefficient (denoted by 'mean') is for spatial pattern correlation of temporal mean structure of geostrophic GMOCs. See **Methods** for the definition of the GMOC index. All correlation coefficient statistics for CDA the 95% significance level.

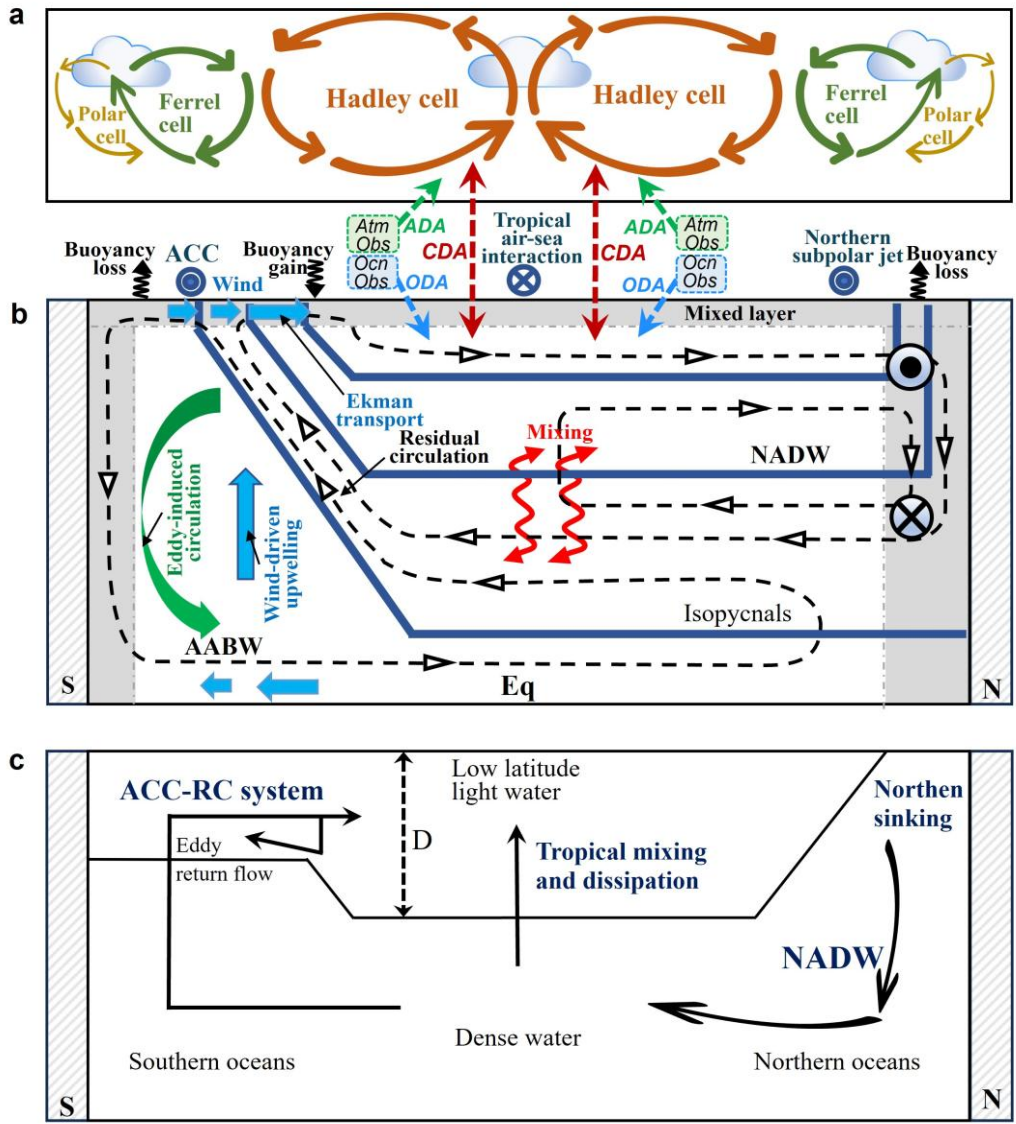

**Extend Data Fig.3 | Schematic illustration of GMOC's integration on cross-sphere climate signals in multiscale processes. a**, Basic atmospheric mean meridional circulation (AMMC) consisting of Hadley, Ferrel and Polar cells. **b**, The configuration of processes at the air-sea interface include [36]: westerly jet at the Antarctic Circumpolar Current (ACC) region, buoyancy gain and loss induced by Ekman effects, northern subpolar jet, upward heat flux and buoyancy loss, and air–sea interactions associated with easterlies in the tropics. The configuration of ocean interior processes include: Northen Hemisphere (NH) sinking to form the North Atlantic Deep Water (NADW), diffusive mixing in tropical oceans, Southern Hemisphere (SH) wind-driven Ekman transport, eddy-induced circulation, and wind-driven upwelling and eddy-induced circulation associated with the Antarctic Bottom Water (AABW). **c**, The mechanism summarized from **b** for GMOC based on the three-term balance model, i.e., the residual between the difference of NH sinking and SH upwelling subtracting tropical upwelling drives the change in pycnocline depth anomaly [2,37].

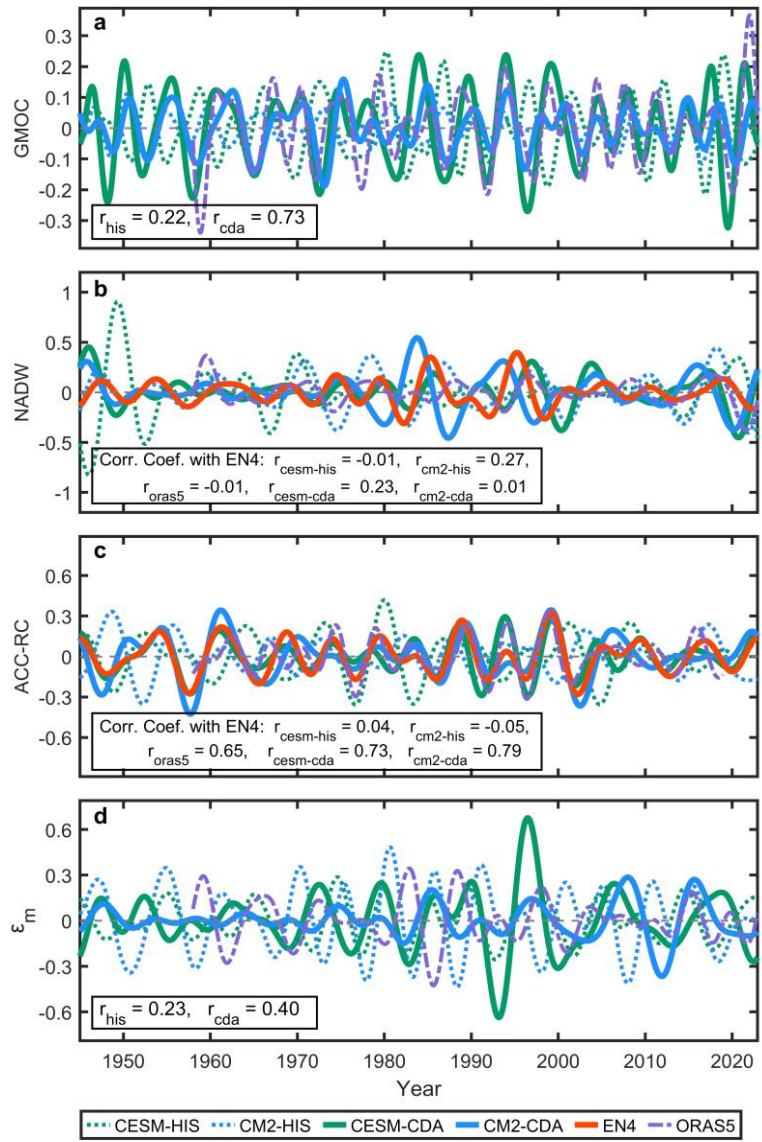

**Extend Data Fig.4 | Observation-consistent interannual variability of GMOC and associated physical modes in coupled reanalyses. a–d**, Same as **Fig.2**, but for 2–5 years filtering of time series for GMOC (**a**), North Atlantic Deep Water (NADW) (**b**), Antarctic Circumpolar Current residual circulation (ACC-RC) (**c**) and tropical diffusive mixing ($\varepsilon_m$) indices (**d**).

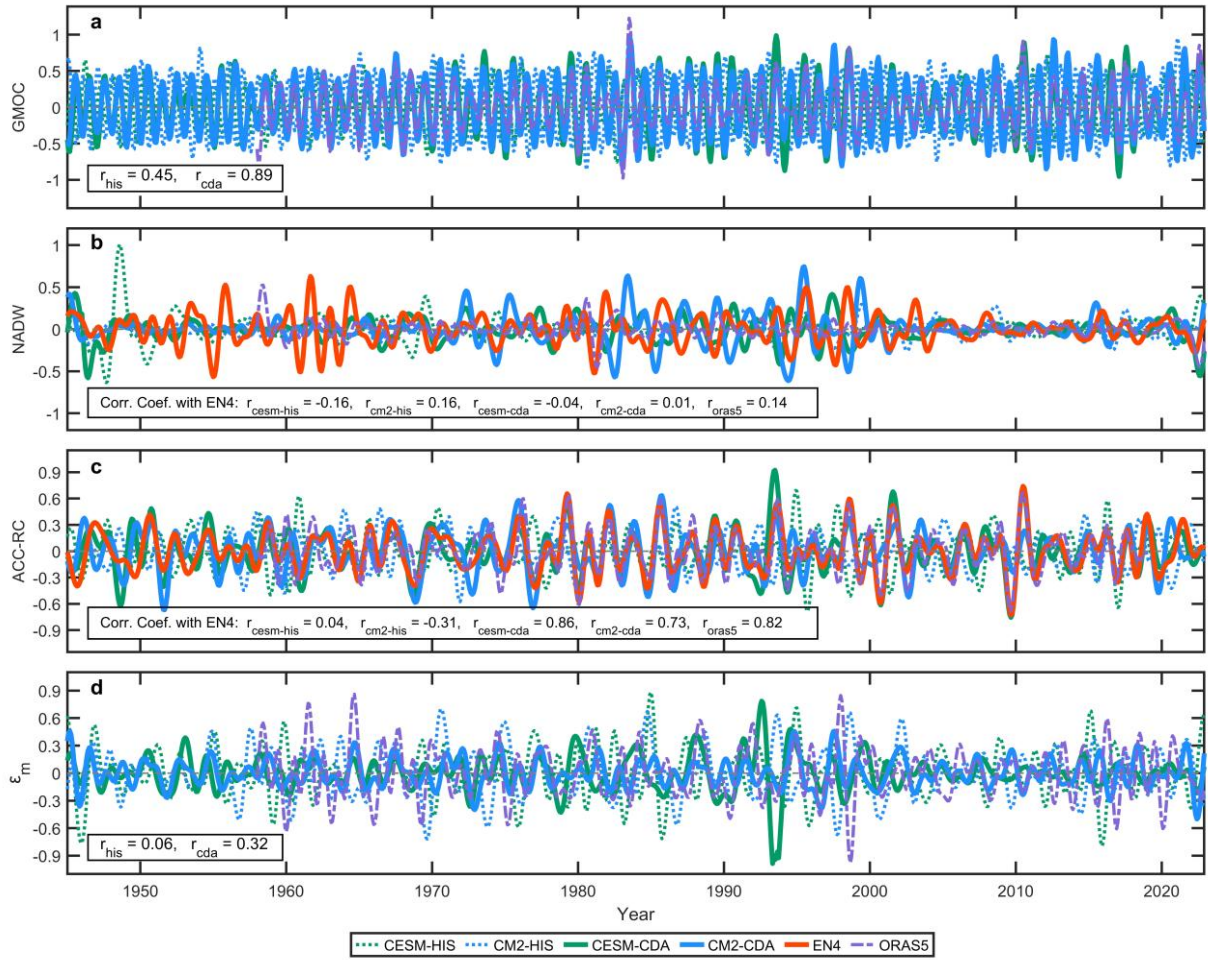

**Extend Data Fig.5 | Observation-consistent seasonal-to-interannual variability of GMOC and associated physical modes in coupled reanalyses. a–d**, Same as **Fig.2**, but for 6 months–2 years filtering of time series for GMOC (**a**), North Atlantic Deep Water (NADW) (**b**), Antarctic Circumpolar Current residual circulation (ACC-RC) (**c**) and tropical diffusive mixing ($\varepsilon_m$) indices (**d**).

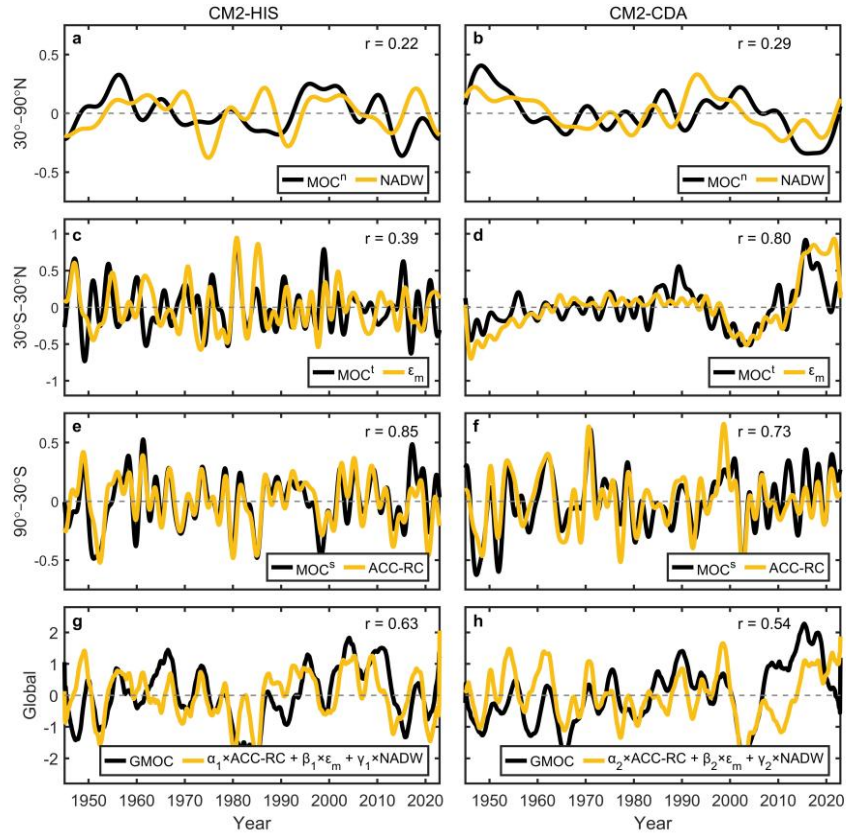

**Extend Data Fig.6 | Composited GMOC with its physical modes in CM2 simulation and coupled reanalysis.** Same as **Fig.3**, but for the model simulation and coupled data assimilation of CM2 (CM2-HIS and CM2-CDA). The parameters α, β, and γ are the linear regression coefficients between the GMOC indices and ACC-RC, $\varepsilon_m$, NADW indices, respectively, as $[\alpha_1, \beta_1, \gamma_1] = [0.39, -0.28, 0.04]$ and $[\alpha_2, \beta_2, \gamma_2] = [0.30, 0.39, 0.01]$.